\documentclass[a4paper,12pt]{article}
\usepackage{amssymb}

\usepackage[utf8]{inputenc}
\usepackage{a4wide}

\usepackage{amsmath}
\usepackage{cite}
\usepackage{xcolor}
\usepackage{amssymb}
\usepackage{amsfonts}
\usepackage{booktabs}
\usepackage{makecell}
\usepackage{multirow}
\usepackage{acronym}
\usepackage{pdflscape}
\usepackage{textcomp}
\usepackage{gensymb}
\usepackage{bigstrut}
\usepackage{array}
\usepackage{enumerate}
\usepackage[small,bf]{caption}
\usepackage{subcaption}
\usepackage{diagbox}
\usepackage{graphicx}
\usepackage{hyperref}
\usepackage{mathbbol}
\usepackage{appendix}
\usepackage{verbatim}
\usepackage{geometry}
\usepackage{makecell}
\usepackage{listings}
\usepackage{mathtools}
\usepackage{dsfont}
\usepackage{longtable}
\usepackage{threeparttable}
\usepackage{floatpag}
\usepackage{arydshln}
\usepackage[capitalise]{cleveref} 
\usepackage{tikz}
\usetikzlibrary{fit,matrix}

\DeclareMathAlphabet{\mathdutchcal}{U}{dutchcal}{m}{n}

\newcommand{\mathsym}[1]{{}}

\newcommand{\qed}{\nobreak \ifvmode \relax \else \ifdim\lastskip<1.5em \hskip-\lastskip \hskip1.5em plus0em minus0.5em \fi \nobreak \vrule height0.75em width0.5em depth0.25em\fi}
\newcommand{\diag}{\mbox{diag}}

\allowdisplaybreaks
\numberwithin{equation}{section}

\DeclareMathOperator{\tr}{Tr}
\DeclareMathOperator{\im}{Im}

\makeatletter
\newcommand*{\org@overidelabel}{}
\let\org@overridelabel\@verridelabel
\@ifpackagelater{acronym}{2015/03/21}{%
  \renewcommand*{\@verridelabel}[1]{%
    \@bsphack
    \protected@write\@auxout{}{\string\AC@undonewlabel{#1@cref}}%
    \org@overridelabel{#1}%
    \@esphack
  }%
}{%
  \renewcommand*{\@verridelabel}[1]{%
    \@bsphack
    \protected@write\@auxout{}{\string\undonewlabel{#1@cref}}%
    \org@overridelabel{#1}%
    \@esphack
  }%
}
\makeatother

\begin{document}

	\begin{titlepage} 
		\begin{center} \hfill \\
		\hfill \\
			\textbf{\Large
Reducing Complex Phases and other \\Subtleties of CP Violation  }

			 \vskip 1cm \vskip 1cm
José Filipe Bastos \footnote{jose.bastos@tecnico.ulisboa.pt} and
J.I. Silva-Marcos\footnote{juca@cftp.tecnico.ulisboa.pt}

\vskip 0.07in Centro de
F{\'\i}sica Te\'orica de Part{\'\i}culas, CFTP, \\ Departamento de
F\'{\i}sica,\\ {\it Instituto Superior T\'ecnico, Universidade de Lisboa, }
\\ {\it Avenida Rovisco Pais nr. 1, 1049-001 Lisboa, Portugal} \end{center}

		\begin{abstract} 

When extending the SM, e.g. with vector-like quarks (VLQ), one inevitably encounters more than one CP violating complex phases. In this work, we develop a method useful in reducing the number of complex phases in scenarios with one and two VLQs. We show that these VLQ-models can be described with just one or two physical complex phases respectively and present an explicit alternative parametrization for the quark mixing matrix. 
Making use of this Reduction of Complex Phases in the case of the Bento-Branco-Parada (BBP) model, i.e. the simplest implementation of the Nelson-Barr solution to the strong CP problem, one also recovers the single phase formulation of BBP model. We comment on how this specific scenario can be distinguished from the general VLQ case with weak-basis invariants.  

\end{abstract}
	\end{titlepage}
	
\section{Introduction}

Although the SM allows for the existence of CP violation, specifically in the flavour structure of the quark sector, it is well known that the predicted CP violation does not meet the Sakharov conditions for baryogenesis \cite{Sakharov:1967dj} and is therefore unable to explain the observed imbalance between matter and anti-matter in the universe. This has motivated a wide variety of extensions of the SM that aim at inducing new and stronger sources of CP violation. In this work we focus on theories that extend the quark sector with vector-like quarks (VLQs). 

In the SM, CP violation (CPV) arises from the complex structure of the Yukawa couplings of the three generations of quarks, generating complex mass matrices and resulting in a complex CKM mixing matrix. This complexity can be parametrized with a single CP violating physical phase. When introducing VLQs to the theory, additional complex couplings are allowed, in general leading to extra physical phases and, potentially, an enhancement of CP violation \cite{Aguilar-Saavedra:2004roc,Botella:2002fr}. VLQs can also provide a framework to solve the strong CP problem, specifically through the Nelson-Barr mechanism \cite{Nelson:1983zb,Nelson:1984hg,Barr:1984qx}, which presents a possible justification for the apparent lack of CPV in strong interactions\footnote{For more recent studies on VLQ of Nelson-Barr type consider \cite{Cherchiglia:2020kut,Cherchiglia:2021vhe,Alves:2023cmw}.}.

Understanding the implications for CPV in the flavour sector of these beyond the SM (BSM) scenarios requires the study of CP-odd weak basis invariant (WBIs) quantities that relate to all these CPV complex phases. The vanishing or not of such quantities can thus indicate the presence of new sources of CPV enhancement. 
However, typically, the number of complex phases will quickly grow with the number of added VLQs, and a complete understanding of the role and importance of these new phases can rapidly prove to be a strenuous task.

In this work we make use of the Reduction of Complex Phases (RCP) technique first introduced in \cite{Silva-Marcos:2002upu} and apply it to the quark sector in the context of extensions of the SM with VLQs. With this technique one can obtain minimal\footnote{We refer to a WB as "minimal" if in that WB, the quark mass matrices can be written in terms of a number of independent parameters that exactly coincides with the number of physical parameters, i.e. the quark masses and the independent parameters parametrizing the quark mixings.} weak bases (WBs) where the number of independent physical phases present in the Yukawa and bare mass couplings is reduced to a minimum, in general lower than the number of physical phases typically obtained by performing the $2n-1$ independent rephasings of the $n$ quark fields of the theory. Here, each physical phase that is explicitly removed from the parametrization of a given WB is implicitly traded for a real parameter or, in the context of the quark mixing, a mixing angle. Hence, when employing this technique one can switch between a minimal WB depending on a minimal number of mixing angles, and a minimal WB depending on a minimal number of CP violating physical phases.
This exchange can then be useful in decluttering the phase content of a given model, in pinpointing which phases can induce new, beyond the SM, sources of CPV and in comparing how different models may enhance CPV. 

Another potential usefulness of RCP might come about in models with more then one standard Higgs in connection with spontaneous CPV.

Ultimately, our goal with this work is to demonstrate how RCP might be employed to promote a better understanding of CPV in BSM scenarios with an extended quark sector.

The paper is organized as follows: in \cref{sec:CPV} we briefly review and compare how CPV is manifested in the SM and in extensions with VLQ singlets, where we emphasise the existence of new sources of CPV in such extensions. In \cref{sec:RCP} we demonstrate the usefulness of the RCP technique in extracting interesting cases where CPV-sensitive weak basis invariants (WBIs) vanish. We also put forward a new parametrization of the quark mixing matrix in the one VLQ singlet case that depends solely on a CP violating phase. In \cref{sec:BBP} we focus on the BBP implementation of the Nelson-Barr solution to the strong CP problem and discuss how a Nelson-Barr VLQ can be distinguished from a generic one in a weak basis invariant way.

\section{CP Violation and VLQs}
\label{sec:CPV}
When considering extensions of the SM, one is confronted with more than one complex CP violating phases. In certain cases this number may indeed be very large. Are all these phases relevant to the same extent? Here in different cases of extensions of the SM with VLQs, we argue that this depends critically on the parametrization one is considering, which may lead to surprising outcomes.

\subsection{CPV in the SM}
As known, in the SM there is only one complex CP violation phase. If one chooses a standard parametrization of the CKM matrix \cite{ParticleDataGroup:2022pth} where 
\begin{equation}
V_\text{CKM}=O_{23}K_{\delta}O_{13}O_{12}
\end{equation}
with the $O_{ij}$ being real orthogonal rotations in $ij-$plane with angle $\theta_{ij}\in[0,\pi/2]$ and where $K_{\delta}=\text{diag}(1,1,e^{i\delta})$ contains the complex phase $\delta\in [0,2\pi]$, then one finds for the rephasing invariant quartet of $V_\text{CKM}$ which expresses CP violation: 
\begin{equation}
J_\text{SM}=|\text{Im}(V_{12}V_{23}V^*_{22}V^*_{13})|=
c_{12}c_{23}c^2_{13}
s_{12}s_{23}s_{13}\left|\sin{\delta}\right|.
\label{Im}
\end{equation}
Another way to characterize this quartet which signalizes CP violaton is by calculating the weak basis invariant 
\begin{equation}
I_\text{SM}=\left|\det([h_u,h_d]^3)\right|=\Delta_0\ J_\text{SM}
\label{w1}
\end{equation}
where we defined the squared Hermitian mass matrices $h_{u,d}=(m\,m^\dagger)_{u,d}$ and the product of quark-mass differences $\Delta_0=6(m^2_t-m^2_c)(m^2_t-m^2_u)(m^2_c-m^2_u)(m^2_b-m^2_s)(m^2_b-m^2_d)(m^2_s-m^2_d)$.

In the SM, all other rephasing invariant quartets (take e.g. $|\text{Im}(V_{11}V_{22}V^*_{12}V^*_{21})|$) give exactly the same answer. However, in extensions of SM with one singlet VLQ, this is no longer the case. In that case, there are more independent rephasing invariant quartets.

In Eqs. (\ref{Im}, \ref{w1}), the obvious condition for CP violation to be absent is when the complex phase $\delta=0$. However, it is worthwhile to stress with some emphasis that this not the only condition. If, for example, $\theta_{13}$ were to vanish then $I_\text{SM}$ would also vanish. However in the SM, with the known experimental constrains this is of course not feasible. 

Nevertheless here, we want to explicitly emphasize, in contrast with the conditions coming from complex phases, the existence of these real parameter-conditions for CP violation absence and that although being non-realistic in the pure SM, these type of real-parameter-conditions may play a role (at least in our perception in the conditions for CP violation) in extensions of SM.

\subsection{CPV in extensions with one down-type VLQ singlet}

One of the simplest extensions of SM is obtained by adding just one singlet VLQ. With respect to CP violation, the resulting effect of this introduction is to increase the number of complex phases significantly. In fact, in a general model with $n_u$ and $n_d$ VLQ isosinglets the number of complex parameters is usually taken to be $1+2(n_u+n_d)$ \cite{Alves:2023ufm}. However as we shall see, depending on the Weak Basis parametrization one chooses, the number of complex phases may be significantly reduced.

Consider as an example the extension of the SM with one (down type) vector-like quark $B$ of electric charge $Q=-1/3$ added to the SM. The Yukawa Lagrangian is given by
\begin{equation}
    -\mathcal{L}_Y=\overline{q}^0_L \phi\  \mathcal{Y}_d \  
        d^0_R + \overline{B}^0_L M_d \ d^0_R  
        +\overline{q}^0_L \Tilde{\phi}\  Y_u\ u^0_R + \text{h.c.},
        \label{lag}
\end{equation}
where $\mathcal{Y}_d$ is a $3\times 4$ matrix with the down Yukawa couplings, $M_d$ a $1\times 4$ matrix of down bare mass terms and the $3\times 3$ matrix $Y_u$ contains the up-quark Yukawa couplings. The $d^0_R$ stands for a vector in generation space with $(d^0_{1R},d^0_{2R},d^0_{3R},B^0_R)$ and likewise $u^0_R$ for $(u^0_{1R},u^0_{2R},u^0_{3R})$. Just as in the SM, we have a standard three doublets
\begin{equation}
    q^0_{iL}=\begin{pmatrix}
        u^0_i\\
        d^0_i
    \end{pmatrix}_{L}.
\end{equation}
After spontaneous symmetry breaking we obtain the mass matrices 
\begin{equation}
    -\mathcal{L}_M=\overline{D}^0_L \mathcal{M}_d D^0_R +\overline{u}^0_L m_u
        u^0_R  + \text{h.c.},
\end{equation}
where we defined
\begin{equation}
    D^0_{L,R}=\begin{pmatrix}
        d^0\\
        B^0
    \end{pmatrix}_{L,R}
\end{equation}
and introduced the quark mass matrices, $m_u$ and $\mathcal{M}_d$, for the up and down sector respectively, given by 
\begin{equation}
   m_u=v Y_u, \quad \quad \quad \mathcal{M}_d=\begin{pmatrix}
        m_d\\
        M_d
    \end{pmatrix}, \hspace{5pt}  m_d=v \mathcal{Y}_d. 
    \label{eq:mu}
\end{equation}
We may choose a weak-basis where the $3\times 3$ up-quark mass matrix $m_u$ is real and diagonal.
The $4\times 4$ down-quark mass matrix $\mathcal{M}_d$ is then diagonalized by the following $4\times 4$ unitary matrix
$\mathcal{V}$, where we choose a standard-like parametrization such that the first row and last ($4$th) column are real\footnote{This parametrization is slightly different but equivalent to the Botella-Chau parametrization \cite{Botella:1985gb}.},
\begin{equation}
   \mathcal{V}\,=\,
\mathcal{O}_{34}\mathcal{O}_{24}\mathcal{O}_{14}\cdot\mathcal{K}_{\alpha \beta}\cdot \mathcal{O}_{23}\mathcal{K}_{\delta}\mathcal{O}_{13}\mathcal{O}_{12},
\label{V}
 \end{equation}
with $\mathcal{K}_{\delta}=\text{diag}(1,1,e^{i\delta},1)$, $\mathcal{K}_{\alpha\beta}=\text{diag}(1,e^{i\alpha},e^{i\beta},1)$, with $\delta,\alpha,\beta\in[0,2\pi]$ and $\theta_{14},\theta_{24},\theta_{34}\in[0,\pi/2]$. For this model, the mixing is then represented by the non-unitary matrix ${V}_\text{CKM}$, the $3\times 4$ matrix which corresponds to the first three rows of $\mathcal{V}$. Thus, in this parametrization of the mixing matrix, apart from three more mixing angles, we have two more complex phases than in the SM.

One way to study CP violation is through the complex rephazing invariants formed by quartets of the mixing matrix $\mathcal{V}$.
These are of particular physical interest as they occur in physical processes described e.g. with box diagrams.
Evaluating these quartets using \cref{V} and defining
\begin{equation}
J_{ijkl}\,\equiv \,\im [\mathcal{V}_{ij}\mathcal{V}_{kl}\mathcal{V}_{il}^*\mathcal{V}_{kj}^*]
\label{JJ}
 \end{equation}one obtains e.g. 
\begin{equation}
\begin{split}
J_{1223}=& \ 
c_{12}c_{23}c_{13}^{2}c_{24}^{2}c_{14}^{2}\ s_{12}s_{23}s_{13}\ \sin \delta 
+c_{12}c_{23}c_{13}c_{24}c_{14}^{2}\ s_{12}s_{13}^{2}s_{24}s_{14}\ \sin \alpha  \\ 
\\ 
& -c_{13}c_{24}c_{14}^{2}s_{12}^{2}s_{23}s_{13}s_{24}s_{14}\
\sin (\alpha +\delta ).
\end{split}
\label{j1223}
\end{equation}
Note however that with SM-like values for the angles $s_{12},s_{23},s_{13}$ and assuming small NP mixings, i.e. $s_{24},s_{14}\lesssim s_{23}$ \cite{Alves:2023ufm}, we find that the first term here is clearly dominant. It roughly corresponds to what one obtains in the SM. Therefore, it is not realistic to have $J_{1223}=0$.
Notwithstanding, if one computes another quartet
\begin{equation}
\begin{array}{l}
J_{1324}=
c_{13}c_{24}c_{14}^{2}\ s_{23}s_{13}s_{24}s_{14}\ \sin(\alpha+\delta ),  
\end{array}
\label{j1324}
\end{equation}
one finds that it vanishes if $\alpha =-\delta$, or even if e.g. $s_{24}=0 $. The same applies to
\begin{equation}
\begin{array}{l}
J_{1334}=
c_{13}c_{24}c_{14}^{2}s_{13}
s_{34}s_{14}\left[  s_{23}s_{34}s_{24}\ \sin (\alpha+\delta ) 
-c_{23}c_{34}
\  \sin (\beta+\delta ) \right]
\end{array}
\label{j1334}
\end{equation}
when $\alpha =$ $\beta=$ $-\delta$ or if the parameter relation between angles and phases $ \ s_{23}s_{34}s_{24}\ \sin (\alpha+\delta ) 
=c_{23}c_{34}
\  \sin (\beta+\delta )$ is satisfied.

Therefore we conclude that certain quantities sensitive to CP violation may be very small or even vanish without clearly contradicting any plausible physical data.

\section{Reduction of Complex Phases (RCP)}
\label{sec:RCP}

Next, we argue that, for some models beyond the SM, the number of CP-violating complex phases depends critically on the choice of WB.

For the model and parametrization of the mixing matrix just described, we have three complex phases. This is also evident when we consider the following minimal weak-basis, where the down-quark matrix is partially diagonal and real but, where the up-down mass matrix, has the following form \cite{Alves:2023ufm}:
\begin{equation}
    \mathcal{M}_d=\begin{pmatrix}
       r_1 & 0& 0 & \Bar{r}_1\\
        0 & r_2 & 0 & \Bar{r}_2\\
        0 & 0 & r_3 & \Bar{r}_3\\
        0 & 0   &  0  & r_4\\
    \end{pmatrix}_d\ ,\  m_u=\begin{pmatrix}
        r_1 & c_1 & c_2\\
        0& r_2 & c_3\\
        0 & 0 & r_3 
    \end{pmatrix}_u
\label{weak1}
\end{equation}
where the $r$ and $c$ are generic labels that represent real and complex parameters, respectively. It is clear that here we have 16 parameters. This corresponds exactly to the same 16 physical parameters: 7 quark masses and 9 parameters in the mixing matrix, i.e. 6 mixing angles and 3 phases.

We now introduce a technique, described in \cite{Silva-Marcos:2002upu}, where at the cost of a real-orthogonal rotation, we are able to eliminate complex phases. This happens when we apply the next real-orthogonal rotation to two complex numbers
\begin{equation}
\begin{pmatrix}
 a_1 + i b_1\\ a_2+i b_2
    \end{pmatrix} =\begin{pmatrix}
 c_\theta & s_\theta \\ -s_\theta & c_\theta
    \end{pmatrix}
    \begin{pmatrix}
 a'_1 \\ a'_2+i b'_2
    \end{pmatrix} 
\label{comp1}
\end{equation} We have thus eliminated the imaginary number $ib_1$. Therefore, two complex complex numbers can be written as one real number and one complex number plus a real-orthogonal rotation, which at the end, amounts to the same number of parameters. In this section, we will apply this method to extensions with one VLQ singlet and to extensions with both one up-type and one down-type VLQ singlet.

In the sequel, we use this procedure to write a new weak-basis for our SM extension with one singlet VLQ to eliminate two of the complex phases in \cref{weak1}. We refer to \cref{ap1} for more details. 

\subsection{Extensions with one down-type singlet VLQ}
\label{sec:1VLQ}
Consider again the case of the SM extended with a down-type singlet. We now use the RCP technique in order to obtain WBs which have a reduced number of phases.

\subsubsection*{Weak Basis A:}

After renaming (for convenience) the parameters used in the appendix, we find the following non-trivial weak basis,
\begin{equation}
    \mathcal{M}_d=\begin{pmatrix}
       r_1 & 0& 0 &r_5 \\
        0 & r_2 & r_3 & r_6 \\
        0 & 0 &  r_4& r_7\\
        0 & 0  & 0   & r_8\\
    \end{pmatrix}_d\ ,\  m_u=\begin{pmatrix}
        r_1  & r_2 & r_3\\
        0 & r_4& r_5 e^{i\alpha}\\
        0 & r_6 & r_7 
    \end{pmatrix}_u
\label{weakd}
\end{equation}
with only one complex parameter leading to just one complex phase.
Note that, we continue to have a total of 16 parameters (equal to the physical parameters): 8 real parameters in the down-quark matrix and in the up-quark mass matrix, 7 real and one complex phase.

For the one VLQ singlet model, in this WB, one salient advantage of the reduction of complex phases is the fact that, if in  \cref{weakd} the phase $\alpha=0$, all sources of CP violation are naturally absent.

Other uses of this RCP mechanism in connection with CP violation are evident when studying certain weak-basis invariants (WBIs). As detailed in \cite{Albergaria:2022zaq}, in these extensions WBIs can be constructed from hermitian matrix built from $m_u$, $m_d$ and $M_d$.
For instance, consider the following WBI:
\begin{equation}
    I_d=\text{Im}\left(\tr\left[ h_u h_d  \overline{h}_d  \right]\right),
    \label{inv}
\end{equation}
where
\begin{equation}
        h_q =m_q m^\dagger_q, \hspace{10mm}
        \overline{h}_d =m_d M^\dagger_d M_d m^\dagger_d,
\end{equation}
where we recall the form in \cref{eq:mu}. In this WB, one obtains
\begin{equation}
\begin{split}
    I_d  = & \ \bigg\{ r_{6_d} \Big[r_{3_u} r_{5_d}
   \left(r_{1_d}^2-r_{2_d}^2-r_{3_d
   }^2\right)-r_{7_d} r_{7_u}
   \left(r_{2_d}^2+r_{3_d}^2-r^2_{4_d}\right)\Big]  \\
    & +   \ r_{3_d} r_{4_d}
   \Big[r_{7_u}
   \left(r_{6_d}^2-r_{7_d}^2\right)-r_{3_u} r_{5_d} r_{7_d}\Big]\bigg\} r_{8_d}^2 r_{5_u}\sin\alpha,
\end{split}
\end{equation}

Clearly, this invariant vanishes not only when $\alpha=0$ (or when its attached real  parameter $r_{5_u}=0$), but also when 
\begin{equation}
 r_{3_u}=r_{7_u}=0   
\end{equation}
This combination is not surprising because in this case we may remove the complex phase by a rephasing of the fields. 
However, what is most surprising is that the invariant also vanishes, e.g. when
\begin{equation}
  r_{5_d}= r_{7_u}= 0.  
\end{equation}

The invariant also vanishes when
\begin{equation}
    \begin{split}
         r_{6_d}=r_{7_d}=0,\\
         r_{3_d}=r_{6_d}=0.\\
    \end{split}
\end{equation}

\subsubsection*{Weak Basis B:}

Following this line of thought, it is possible to find more intriguing cases when considering other weak-bases. For instance, in the following weak-basis which may be derived just as well from the general basis (and which as in case A, has only one complex phase), we find
\begin{equation}
    \mathcal{M}_d=\begin{pmatrix}
       r_1 & r_2& r_3 &r_4 \\
        0 & r_5 & r_6 & r_7 \\
        0 & r_8&  r_9\, e^{i\alpha}& r_{10}\\
        0 & 0  & 0   & r_{11}\\
    \end{pmatrix}_d\ ,\  M_u=\begin{pmatrix}
        r_1  & 0 & 0\\
        0 & r_2& r_4 \\
        0 & 0 & r_3 
    \end{pmatrix}_u
\label{weakdo}
\end{equation}
In this basis, the invariant $I_d$ given in \cref{inv} further vanishes also when e.g.
\begin{equation}
  r_{{10}_d}= r_{4_u}= 0,
  \label{r10}
\end{equation}
thus evidencing another structure where certain CP-odd WBIs may disappear. In \cref{app:numerical_ex} we present a realistic numerical example of such a case.

It is important to notice that the vanishing of just one CP-odd WBI is not sufficient to have no CP violation at all. To accomplish this, other CP-odd WBIs must also vanish. Thus even if the conditions in \cref{r10} are satisfied leading to $I_d=0$, then still we have a non-vanishing value for
\begin{equation}
I_0=\text{Im} \left(\tr\left[ h_d , h_u \right]^3\right)=\\
\Delta \sin\alpha,
\label{eq:I0}
\end{equation} 
with 
\begin{equation}
\begin{split}
    \Delta  = & \ 6 \left(r_{3_u}^2-r_{2_u}^2\right)\left(r_{3_u}^2-r_{1_u}^2\right)\left(r_{2_u}^2-r_{1_u}^2\right) r_{8_d}r_{9_d} \\
    & \  \cdot \left(r_{3_d}r_{5_d}-r_{2_d}r_{6_d}\right)\left({r_2}_d {r_5}_d + {r_3}_d {r_6}_d + {r_4}_d {r_7}_d\right).
\end{split}
\end{equation}

The eminent benefit of this RCP approach is that all CP-odd WBIs are proportional to just one complex phase via $\sin\alpha$. This may enlighten the conditions for having or not having new sources of CP-violation.

Finally, we note that this possibility of having a reduced number of complex phases in only a limited quantity of elements of the mass matrices might be of interest when considering models which include more than one standard Higgs. For example, the presence of a Higgs scalar, with a CP violating VEV, in the $(3,3)$ mass matrix element as in the Weak-Basis of Eq. (\ref{weakdo}) might occur as the result of e.g. a discrete symmetry\footnote{See possible related examples in \cite{Emmanuel-Costa:2017bti}.}.

\subsection{Single phase for the $4\times 4$ Mixing}
\label{sec:param}

In view of the previous reduction of phases, where we found WBs with just one phase as e.g. in Eq. (\ref{weakdo}), one might rightfully ask if this singular reduction is also viable for the standard $4\times 4$ mixing matrix given in Eq. (\ref{V}) which, we emphasize, is usually parametrized with three complex phases.

What we find is that this indeed possible in the region of parameter-space within the limits set by experiment on the CKM mixing angles and CP violation. For details see Appendix \ref{ap2}. 

In this region, instead of the parametrization given in Eq. (\ref{V}) with three phases, it is possible to write the unitary $4\times 4$ matrix $\mathcal{V}$ responsible for the quark mixing as
\begin{equation}
   \mathcal{V}\,=\,
\mathcal{O}_{34}\mathcal{O}_{24}\mathcal{O}_{14} \cdot\mathcal{O}_{23}\cdot\mathcal{O}'_{13}\mathcal{O}'_{12}\cdot\mathcal{K}_{\sigma}\mathcal{O}_{13}\mathcal{O}_{12},
\label{Vu}
 \end{equation}
with only one phase expressed in $\mathcal{K}_{\sigma}=\diag(1,1,e^{i\sigma},1)$.

Subsequently, all quantities which indicate CP violation will be proportional to the only complex phase via $\sin{\sigma}$. For example, computing as in Eq. (\ref{j1324}), $J_{1324}$, we find \begin{equation}
\begin{array}{l}
J_{1324}=
c_{13}c_{24}c_{14}^{2}\ s_{13}s_{24}s_{14}\ (c'_{23}s_{23}+c_{23}s'_{12}s'_{13}) \sin{\sigma}.
\end{array}
\label{oj1324}
\end{equation}

\subsection{Extensions with up and down-type singlet VLQs}
Let us now consider the case of the SM with two  vector-like quarks, a down-type and an up-type singlet VLQ. The Lagrangian for this case \cite{Alves:2023ufm} is the obvious extension of the one given in \cref{lag}.

It is easy to find a weak-basis which reflects the $22$ physical parameters for this case of 2-VLQ-up-down singlets,
\begin{equation}
    \mathcal{M}_d=\begin{pmatrix}
       r_1 & 0& 0 & 0\\
        0 & r_2 & 0 &0 \\
        0 & 0 & r_3 & 0\\
        \Bar{r}_1& \Bar{r}_2   & \Bar{r}_3  & r_4\\
    \end{pmatrix}_d, \hspace{10mm}      
\mathcal{M}_u=\begin{pmatrix}
        r_1 & r_2 & r_3& r_4\\
        0& c_1 & c_2&c_3 \\
        0 & 0 &c_4  & c_5\\
         0 & 0 & 0 & r_5\\
    \end{pmatrix}_u,
\label{weak1o}
\end{equation}
here having a total of five complex phases.

However, after using our method of the reduction of complex phases, it is possible to arrive at a weak-basis with only two phases:
\begin{equation}
    \mathcal{M}_d=\begin{pmatrix}
       r_1 & 0& 0 & 0\\
        0 & r_2 & r_4 &0 \\
        0 & r_5 & r_3 & 0\\
        r_6& r_7   & r_8 \, e^{i\alpha} & r_9\\
    \end{pmatrix}_d, \hspace{10mm}     
\mathcal{M}_u=\begin{pmatrix}
        r_1 & r_4 & r_7& r_9\\
        0& r_2 & r_5&r_8 \\
        0 & r_6 &r_3 \, e^{i\beta}  & r_{10}\\
         0 & 0 & 0 & r_{11}\\
    \end{pmatrix}_u.
\label{weak1a}
\end{equation}
For this case we can build extra CP-odd WBIs, such as
\begin{equation}
I_u=\im\left(\tr\left[ h_u h_d \overline{h}_u \right]\right)=P_u\,r_{11_u}^2\,r_{3_u}\,\sin \beta 
\label{invu0}
\end{equation}
where $P_u$ is a polynomial in the $r$'s of the up-down sectors, but which reduces to 
\begin{equation}
P_u=\left[ \left(r_{3_d}^2-r_{1_d}^2\right)r_{7_u}r_{9_u}+\left(r_{3_d}^2-r_{2_d}^2\right)r_{5_u}r_{8_u}
\right]r_{10_u} 
\label{invu}
\end{equation}
when $r_{4_d}=r_{5_d}=0$.

In addition, and in order to obtain a CP-odd WBI which is only dependent on the other complex phase $\alpha$ in $\mathcal{M}_d$, we may evaluate 
\begin{equation}
I'_d=\im\left(\tr\left[ h_d \overline{h}_u \overline{h}_d \right]\right)=P_d\,r_{11_u}^2\,r_{8_d}\,\sin\alpha 
\label{invd}
\end{equation}
where again $P_d$ is a polynomial in the $r$'s, but which with the previous conditions $r_{4_d}=r_{5_d}=0$, reduces to 
\begin{equation}
P_d=\left[ (r_{3_d}^2-r_{1_d}^2)r_{6_d}r_{1_d}r_{9_u}+(r_{3_d}^2-r_{2_d}^2)r_{7_d}r_{2_d}r_{8_u}
\right]r_{10_u}\,r_{3_d}, 
\label{invdo}
\end{equation}

We find that $I_u$ and $I'_d $ do not only vanish when the obvious $r_{3_u}\sin \beta=0$ respectively $r_{8_d}\sin\alpha=0$. Under the previous restrictions (i.e. $r_{4_d}=r_{5_d}=0$), surprisingly they also vanish when $r_{10_u}=0$. 
However even then, under all these combined conditions i.e.  $r_{4_d}=r_{5_d}=r_{10_u}=0$, we still obtain for the other CP-odd WBI $I_0$ in \cref{eq:I0} a non-vanishing value, 
\begin{equation}
I_0=
\Delta' \sin\beta,
\label{Jo}
\end{equation} 
where 
\begin{equation}
\begin{split}
\Delta' = & \ 6\ (r_{3_d}^2-r_{2_d}^2)(r_{3_d}^2-r_{1_d}^2)(r_{2_d}^2-r_{1_d}^2) (r_{4_u}r_{5_u}-r_{2_u}r_{7_u})\\
& \quad \cdot r_{3_u} r_{6_u}
(r_{2_u}r_{4_u}+r_{5_u}r_{7_u}+
r_{8_u}r_{9_u}). 
\end{split}
\end{equation}

Finally and in a similar way, one may also significantly reduce the number of complex CP violating phases (RCP) in other models, either with more singlet-VLQs or with doublet-VLQs.

\section{The BBP-Nelson-Barr-type Model}
\label{sec:BBP}

Next, and in connection with our RCP mechanism, we focus on the Bento-Branco-Parada (BBP) model \cite{Bento:1991ez}, the simplest realisation of the Nelson-Barr proposal. 
As known, the Nelson-Barr solution to the strong CP problem is based on spontaneously broken CP with the strong CP parameter $\Bar\theta$ naturally vanishing at tree level.

The BBP model has many of the features given in the Lagrangian of \cref{lag} as it also involves a down-type singlet VLQ. In addition it includes a scalar singlet $S$ which couples to the VLQ. These new fields transform non-trivially under a simple $Z_2$ symmetry. Before the scalars obtain vacuum expectation values (VEVs), CP conservation is assumed, meaning that the matrices $\mathcal{Y}_d$ and $Y_u$ are real. The fourth column of $\mathcal{Y}_d$ is $0$ due the $Z_2$ symmetry.
The $1\times  4$ matrix $M_d$ is now replaced by (real) couplings $y_i,y'_i$ to the $S$ scalar and a (real) bare mass term $M$, i.e. $M_d=(y_i S+y'_i S^*, M)$.
Finally and due to spontaneously broken CP and a complex VEV of the singlet scalar, one obtains an up quark mass matrix $m_u$ which is real and the a down quark mass matrix with the following form
\begin{equation}
    \mathcal{M}_d=\begin{pmatrix}
       v \hspace{2pt} Y_d & 0\\
        C  & M\\
    \end{pmatrix}
\label{massbbp}
\end{equation}
where the $3\times 3$ matrix $Y_d$ (the first three columns of $\mathcal{Y}_d$) contains real Yukawa couplings of SM quarks, whereas $C$ contains the complex couplings of VLQs to right-handed SM quarks, i.e. 
\begin{equation}
    C=(r_1 \hspace{2pt} e^{i\alpha_1},r_2 \hspace{2pt} e^{i\alpha_2},r_3 \hspace{2pt} e^{i\alpha_3}).
\end{equation}
The crucial ingredient here is that, explicitly one has that the $\arg[\det( \mathcal{M}_d)]=0$, thus leading to a simple solution to the strong CP problem.

One may now ask whether the BBP mass matrix structure given in \cref{massbbp}  might, in some way, be obtained from a general weak-basis (involving as in the BBP model only one down-VLQ) with our RCP mechanism using only real orthogonal transformation matrices. The latter do not contribute to the $\Bar\theta$ parameter. 
If this were possible and having complex Yukawa couplings $\mathcal{Y}_d$, then in some sense BBP-like models might be related to the general case through a choice of a weak-basis.
Fortunately, the answer to this question is a definite: no, this is not possible! The reason for this is linked to concept of WBIs.

A distinctive feature of the BBP model is that for the weak-basis invariants $I_d$ defined in \cref{inv} and $I_0$ in \cref{eq:I0}, we have 
\begin{equation}
 I_d\neq 0\  , \ \ \ I_0= 0  \ \ \ \ \ \ {{\rm (BBP \  model)}}.
  \label{bbpinv}
\end{equation}
Thus, CP violation in the BBP model is signalled by the $I_d$ and not by the $I_0$, which must vanish.

Conversely, if for some VLQ-like model $I_0\neq 0$ because there is some complex phase in the matrix $m_d$ (assuming as in BBP that the up-quark mass matrix $m_u$ is real), then there is no weak-basis transformation which will change this, precisely because $I_0$ is a WBI, thus answering the question.

For singlet VLQ models, one might expect that the invariant $I_0$ of \cref{eq:I0} to be somewhat related to the standard model $I_\text{SM}$ of \cref{w1}, and indeed that is the case in some models. However as it is clear, this is definitely not case for the BBP model. To arrive at a comparative  $I_\text{SM}$-like WBI, one must construct an effective Hermitian squared down mass matrix $h^\text{eff}_d$, which for \cref{massbbp} takes the form
\begin{equation}
    h^\text{eff}_d=v^2 Y_d \left(\mathbb{1}_3-{\frac{C^\dagger C}{(CC^\dagger+M^2)}}\right) Y_d^\dagger,
\label{eff}
\end{equation}
so that $h^\text{eff}_d$,
contrary to $h_d$ and $h_u$,
is complex, and one may then calculate
\begin{equation}
I_\text{eff}=\tr([h_u,h^\text{eff}_d]^3)
\label{weff}
\end{equation}
which will signal CP violation. The effect of $h^\text{eff}_d$ is to generate the SM-effective CP violation.

Using our RCP procedure it is of course also possible to reduce the number of the complex phases of the BBP model in Eq. (\ref{massbbp}) to just one phase without affecting  $\Bar\theta$, thus arriving at the following weak-basis
\begin{equation}
    \mathcal{M}_d=\begin{pmatrix}
       r_1 &  r_2 &r_3 &0\\
       0 &  r_4 & r_5 & 0 \\
       0 &  r_6 & r_7 & 0\\
       r_8 & r_9 & r_{10} \,e^{i\alpha}  & M\\
    \end{pmatrix}_d , \hspace{5mm} M_u=\begin{pmatrix}
        r_1  & 0 & 0\\
        0 & r_2& 0\\
        0 & 0 & r_3 
    \end{pmatrix}_u,
\label{massbbpn}
\end{equation}
where the number of physical parameters is reduced to 15, which is one less than the generic case.

A distinction between the BBP and the generic VLQ scenarios in terms of the total number of parameters and the number of complex parameters had already been established in \cite{Cherchiglia:2020kut}.
However, as we have shown here, the generic one-VLQ case can also be described in terms of a single phase.

In the limit of extremely high energies, the so called Extreme Chiral Limit (ECL) where to a good approximation $m_u=m_c=m_d=m_s=0$, after WB-transformations, one arrives in the BBP model at mass matrices of the form
\begin{equation}
    \mathcal{M}_d=\begin{pmatrix}
       0 &  0 & 0 &0\\
       0 &  0 & 0 & 0 \\
       0 &  0 & r_7 & 0\\
       0 & 0 & r_{10} \,e^{i\alpha}  & M\\
    \end{pmatrix}_d , \hspace{5mm} M_u=\begin{pmatrix}
        0  & 0 & 0\\
        0 & 0 & 0\\
        0 & 0 & r_3 
    \end{pmatrix}_u,
\label{massbbpn}
\end{equation}
However, even here one finds that the phase $\alpha$ is non-physical as it can be removed with a rephasing of the fields $B^0_{L,R}\rightarrow e^{i\alpha} B^0_{L,R}$. This, of course, implies that in this limit, and contrary to the general VLQ singlet case studied in \cite{Albergaria:2022zaq,delAguila:1997vn}, there is no CPV at extremely high energies.

\section{Conclusions}

In this work we draw attention to the fact that the number of complex parameters relevant for CP violation in BSM models depends critically on the Weak-Basis one is working with. In particular in SM-extensions with just one or two VLQs, we show that, when using the technique of Reduction of Complex Phases, one obtains WBs which have a smaller number of physical CP violating phases than the number one typically obtains from a general naive parameter counting. Subsequently and making use of these new minimal WBs (i.e. with minimal complex parameters), we obtained simpler expressions for some CP-odd WBIs. For the one VLQ singlet case, each CP-odd WBI will depend on just one single physical phase, and in the two VLQ singlet case (one in each sector) we will have just two physical phases. 
In the end, we hope that the question of whether new CPV sources might be relevant or not, is at least partially answered when describing models in terms of less complex parameters.

With regard to Nelson-Barr type models and VLQs, we show that in the BBP scenario with just one VLQ, the number of complex parameters can be the same as the general VLQ case where $\Bar{\theta}$ is overlooked. However, we have found that a distinction between these two scenarios can be established making use of WBIs. We conclude that in the limit of extremely high energies (where the masses of the first two generations can be neglected), in the BBP scenario there is no CP violation, contrary to what happens in the generic VLQ case.

Finally, we emphasise that our method of reducing complex phases might be very useful in identifying mass matrix structures which lead to simple expressions for Weak-Basis invariants linked to experimentally measurable physical quantities.
Our minimally complex WBs may also be useful in models with a larger Higgs content in connection with spontaneous CPV.

\appendix\newpage

\section{Reduction of Complex Phases - RCP}
\label{ap1}
Consider the case of the extension of SM with one VLQ singlet. Other cases are similar, however with their particular constraints.

Starting with our weak-basis in Eq. (\ref{weak1}) we use a real-orthogonal $O_{12}$ rotation in the $(1,2)-$plane of the left-handed fields (i.e. the first two rows), with an angle $\theta$, to turn one complex number $c_2$ into a real number $r'_2$, in the up-quark mass matrix $M_u$. We obtain
\begin{equation}
    \mathcal{M}_d=\begin{pmatrix}
      c_\theta &   -s_\theta  &0& 0\\
         s_\theta  &  c_\theta  & 0 & 0\\
        0 & 0 & 1 & 0\\
       0 &  0 &  0  & 1\\
    \end{pmatrix}\begin{pmatrix}
       r_1 & 0& 0 &  \Bar{r}_1\\
        0 & r_2 & 0 &  \Bar{r}_2\\
        0 & 0 & r_3 & \Bar{r}_3\\
       0&  0 &   0 & r_4\\
    \end{pmatrix}_d\ ,\  m_u=\begin{pmatrix}
        r_1 c_\theta & r'_1 & r'_2\\
        r_1 s_\theta & c'_2 & c'_3\\
        0 & 0 & r_3 
    \end{pmatrix}_u
\label{weaka1}
\end{equation}
where, in addition with a right-handed rephasing of the right-handed up quark field in the second row, we have also obtained an extra real parameter ${r'_1}_u$. Next, we apply to $m_u$ a real-orthogonal $O_{23}$ rotation of the right-handed up-fields to obtain an extra real parameter :
    \begin{equation}
    \mathcal{M}_d=\begin{pmatrix}
      c_\theta &   -s_\theta  &0& 0\\
         s_\theta  &  c_\theta  & 0 & 0\\
        0 & 0 & 1 & 0\\
       0 &  0 &  0  & 1\\
    \end{pmatrix}\begin{pmatrix}
       r_1 & 0& 0 &  \Bar{r}_1\\
        0 & r_2 & 0 &  \Bar{r}_2\\
        0 & 0 & r_3 & \Bar{r}_3\\
       0&  0 &   0 & r_4\\
    \end{pmatrix}_d\ ,\  m_u=\begin{pmatrix}
        r_1 c_\theta & r''_1 & r''_2\\
        r_1 s_\theta & r'''_2 & c''_3\\
        0 & r''_3 & r'_3 
    \end{pmatrix}_u
\label{weakb1}
\end{equation}
to obtain a real parameter in the $(2,2)$ element of the up-quark mass matrix $m_u$.
We do not stop here, because the down and up mass matrices have the $\theta$-angle in common. We will now eliminate this "interconnection", by rotating again the left-handed fields back with the inverse $O^T_{12}$ . Thus, obtaining
   \begin{equation}
    \mathcal{M}_d=\begin{pmatrix}
       r_1 & 0& 0 &  \Bar{r}_1\\
        0 & r_2 & 0 &  \Bar{r}_2\\
        0 & 0 & r_3 & \Bar{r}_3\\
       0&  0 &   0 & r_4\\
    \end{pmatrix}_d\ ,\  m_u=\begin{pmatrix}
        r_1  & \Bar{r}_1 & \Bar{r}_2\\
        0 & \Bar{r}'_2 & \Bar{c}'_3\\
        0 & r''_3 & \Bar{c}''_3 
    \end{pmatrix}_u
\label{weakb2}
\end{equation}where, with a right-handed rephasing of the third row of the up-matrix, we have changed a complex entry originating from the 'back'-rotation $O^T_{12}$ to a real parameter in the $(1,3)$ element of the up-quark mass matrix. However now we obtain a complex parameter in the $(3,3)$ entry.

Finally, applying a real-orthogonal $O_{23}$ rotation,  of the left-handed fields, we arrive at 
   \begin{equation}
    \mathcal{M}_d=\begin{pmatrix}
       r_1 & 0& 0 &\Bar{r}_1 \\
        0 & r'_2 & r''_2 &\Bar{r}'_2 \\
        0 & 0 & r'_3 &\Bar{r}'_3  \\
         0&   0 &   0& r_4\\
    \end{pmatrix}_d\ ,\  m_u=\begin{pmatrix}
        r_1  & \Bar{r}_1 & \Bar{r}_2\\
        0 & \Bar{r}''_2& \Bar{c}\\
        0 & \Bar{r}'_3 & \Bar{r}_3 
    \end{pmatrix}_u
\label{weakb2}
\end{equation}
where, in the down quark mass matrix, we have also rotated the second and third down right-handed fields (expressed here in 2nd and 3rd columns) with a real orthogonal matrix to obtain a zero entry in the $(3,2)$ entry.

We have thus obtained, with WB transformations, a WB where we have 8 real parameters in the down-quark matrix, and 6 real and just one complex parameter in the up-quark matrix, thus a total of 16 parameters equal to our initial number of physical parameters.

\section{$4\times 4$ unitary Mixing with only one phase}
\label{ap2}
Next, we check that the $4\times 4$ unitary mixing in Eq. (\ref{V}), which has three complex phases can be written with only one complex phase. More specifically, we show that it is possible to write $\mathcal{V}_3$ the effective $3\times 3$ part-matrix of $\mathcal{V}$ having the three complex phases, i.e.
\begin{equation}\mathcal{V}_3\equiv\text{diag}(1,e^{i\alpha},e^{i\beta},1)\cdot \mathcal{O}_{23}\cdot\text{diag}(1,1,e^{i\delta},1)\cdot\mathcal{O}_{13}\mathcal{O}_{12}
\label{Voa}
 \end{equation}
 as 
\begin{equation}
 \mathcal{V}_3=
\tilde{\mathcal{O}}_{23}\cdot\tilde{\mathcal{O}}'_{13}\tilde{\mathcal{O}}'_{12}\cdot\text{diag}(1,1,e^{i\sigma},1)\cdot\tilde{\mathcal{O}}_{13}\tilde{\mathcal{O}}_{12}\cdot\mathcal{K}_F
\label{Va}
 \end{equation}
where $\mathcal{K}_F$ is a diagonal 
unitary matrix of unphysical phases which can be incorporated in a redefinition of the quark fields.

To do this, it is easy to check that on the one hand, one may always write every symmetrical $3\times 3$ unitary matrix (and in particular also $\mathcal{V}_3^T \ \mathcal{V}_3$) as
\begin{equation}
\mathcal{V}_3^T \ \mathcal{V}_3=\mathcal{K}_G\cdot
\hat{\mathcal{O}}_{23}^T\hat{\mathcal{O}}_{12}^T\cdot
\mathcal{K}^{(-)}_{2\rho}\cdot\hat{\mathcal{O}}_{12}
\hat{\mathcal{O}}_{23}\cdot\mathcal{K}_G
\label{Vb}
\end{equation} where $\mathcal{K}^{(-)}_{2\rho}=\diag
(1,-1, e^{2i\rho},1)$ and $\mathcal{K}_G$ a  diagonal 
unitary matrix. But, on the other hand, it is also possible to write the same $\mathcal{V}_3^T \mathcal{V}_3$ as 
\begin{equation}
\mathcal{V}_3^T \ \mathcal{V}_3=\mathcal{K}'_G\cdot
\hat{\mathcal{O}}_{23}^T\hat{\mathcal{O}}_{12}^{'T}\cdot
\mathcal{K}_{2\sigma}\cdot\hat{\mathcal{O}}'_{12}
\hat{\mathcal{O}}_{23}\cdot\mathcal{K}'_G
\label{Vc}
\end{equation}
where now the complex phase is a different position with $\mathcal{K}_{2\sigma}=\diag
(1, e^{2i\sigma},1, 1)$ and where $\mathcal{K}'_G$ is also a diagonal 
unitary matrix. In both these equations, the $\hat{\mathcal{O}}_{23}$ is the same, which makes this correspondence more easy to unravel.

However the point here is that, from \ref{Vc} it also follows that one may write
\begin{equation}
\mathcal{V}_3=\hat{\mathcal{O}}\cdot
\mathcal{K}_{\sigma}\cdot\hat{\mathcal{O}}'_{12}
\hat{\mathcal{O}}_{23}\cdot\mathcal{K}'_G
\label{Vd}
\end{equation}
with $\mathcal{K}_{\sigma}=\diag
(1, e^{i\sigma},1, 1)$ and $\hat{\mathcal{O}}$ being an orthogonal real matrix. Then, by using permutations, redefining angles and phases, one arrives easily at the form of \ref{Va},
\begin{equation}
 \mathcal{V}_3=
\tilde{\mathcal{O}}_{23}\cdot\tilde{\mathcal{O}}'_{13}\tilde{\mathcal{O}}'_{12}\cdot\text{diag}(1,1,e^{i\sigma},1)\cdot\tilde{\mathcal{O}}_{13}\tilde{\mathcal{O}}_{12}\cdot\mathcal{K}_F
\label{Vdo}
\end{equation}
 
We have checked that this re-parametrization in terms of mostly angles and only one phase is possible for almost all values in parameter-space. In particular, inserting in \ref{Voa} values, near to the experimental CKM data, for the angles $\theta_{12}\approx 0.225$, $\theta_{23}\approx 0.04$, $\theta_{13}\approx 0.004$ and phase $\delta\approx 1.2$ together with any value for the other phases
$\alpha,\beta$, we find that writing  $\mathcal{V}_3$ in the form of \ref{Va}, is always true.

\section{Numerical example with one Complex Phase}
\label{app:numerical_ex}

Here, we give a realistic numerical example with only one complex phase, in the case of the SM with one singlet VLQ.
In the following example based on Weak-Basis B, CP violation is evident from the $I_0$ defined in \cref{eq:I0}. However for the other CP violation sensitive $I_d$ defined in Eq. (\ref{inv}), we have $I_d=0$.

As explained, the up-quark mass matrix $m_u$ is diagonal and contains the up-quark masses. For the down-quark mass matrix we propose (in GeV)
\begin{equation}
\mathcal{M}_d=\left(
\begin{array}{cccc}
 0.0027 & 0.01722 & 0.00988 & 50 \\
 0 & 0.041 & 0.13 & 6 \\
 0 & 1.3 & -2.6 i & 0 \\
 0 & 0 & 0. & 1218 \\
\end{array}
\right)
\end{equation}
which results in the following masses (in GeV)
\begin{equation}
m_d=0.00263\,,\ m_s=0.0705\,,\ m_b=2.91\,,\ m_B=1219
\end{equation}
$\mathcal{M}_d$ is diagonalized by the following unitary matrix $\mathcal{V}$, with (as explained) the CKM mixing matrix $V_{CKM}$ being the $3\times 4 $ upper-part of  $\mathcal{V}$:
\begin{equation}
|\mathcal{V}|=
\left(
\begin{array}{cccc}
 0.97303 & 0.22696 & 0.004016920 & 0.041016 \\
 0.22715& 0.97300& 0.04048183 & 0.0049219 \\
 0.0089401 & 0.039687& 0.99917 &1.994\times{10}^{-9}
   \\
 0.039116 & 0.013279& 0.00035025 & 0.99914 \\
\end{array}
\right)
\end{equation}
One obtains for this numerical example a value for the $J_{1223}$, as defined in Eq. (\ref{JJ}), of
\begin{equation}
    J_{1223}=3.45\times 10^{-5}.
\end{equation}

In this example, the Cabibbo Angle Anomaly is also addressed by having a 
\begin{equation}
  |\mathcal{V}_{14}|=0.041.
\end{equation}

\providecommand{\noopsort}[1]{}\providecommand{\singleletter}[1]{#1}%

\providecommand{\href}[2]{#2}\begingroup\raggedright

	\end{document}